\documentclass[aps,prb,superscriptaddress,showpacs,twocolumn]{revtex4}
\usepackage{graphicx}
\begin{document}
\title{Nonlocal vs local vortex dynamics in the transversal flux transformer effect}

\author{Florian Otto}
\altaffiliation{Present address: attocube systems AG, Germany.}
\affiliation{Institute for Experimental and Applied Physics, University of Regensburg, D-93025 Regensburg, Germany}
\author{Ante Bilu\v{s}i\'{c}}
\affiliation{Institute for Experimental and Applied Physics, University of Regensburg, D-93025 Regensburg, Germany}
\affiliation{Faculty of Natural Sciences, University of Split, N. Tesle 12, HR-21000 Split, Croatia}
\author{Dinko Babi\'{c}}
\affiliation{Department of Physics, Faculty of Science, University
of Zagreb, Bijeni\v{c}ka 32, HR-10000 Zagreb, Croatia}
\author{Denis Yu.~Vodolazov}
\affiliation{Institute for Physics of Microstructures, Russian
Academy of Sciences, 603950, Nizhny Novgorod, GSP-105, Russia}
\author{Christoph S\"{u}rgers}
\affiliation{Karlsruhe Institute of Technology, Physikalisches Institut and Center
for Functional Nanostructures, D-76128 Karlsruhe, Germany}
\author{Christoph Strunk}
\affiliation{Institute for Experimental and Applied Physics, University of Regensburg, D-93025 Regensburg, Germany}
\begin{abstract}

In this follow-up to our recent Letter [F. Otto {\it et al.}, Phys.~Rev.~Lett. {\bf 104}, 027005 (2010)], we
present a more detailed account of the superconducting transversal flux transformer effect (TFTE) in amorphous ($a$-)NbGe nanostructures in the regime of strong nonequilibrium in local vortex motion. Emphasis is put on the relation between the TFTE and local vortex dynamics, as the former turns out to be a reliable tool for determining the microscopic mechanisms behind the latter. By this method, a progression from electron heating at low temperatures $T$ to the Larkin-Ovchinnikov effect close to the transition temperature $T_c$ is traced over a range $0.26 \leq T/T_c \leq 0.95$. This is represented by a number of relevant parameters such as the vortex transport entropy related to the Nernst-like effect at low $T$, and a nonequilibrium magnetization enhancement close to $T_c$. At intermediate $T$, the Larkin-Ovchinnikov effect is at high currents modified by electron heating, which is clearly observed only in the TFTE.

\end{abstract}
%
\maketitle

\section{Introduction}
\label{introduction}

Applying a transport current $I$ to a type II superconductor in the mixed state may result in vortex motion and power dissipation if the driving force $f_\mathrm{dr}$ on vortices (per unit vortex length $d$) exceeds the pinning force. For a homogeneous mixed state, $f_\mathrm{dr}$ is given by the Lorentz force
$f_\mathrm{L} = j \phi_0$, where $j$ is the transport current density and $\phi_0$ the magnetic flux quantum. When effects related to $j$ leave thermodynamics of the mixed state unchanged, which happens at low $j$,
any nonlinearity in the voltage ($V$) vs $I$ curves  is caused by a competition between $f_\mathrm{L}$ and the pinning force. Further increase of $j$ not only enhances $f_\mathrm{L}$ but can also change the thermodynamic properties if $j$ becomes large enough.\cite{lo,kunchur} Such a strong nonequilibrium (SNEQ) corresponds to a mixed state that is distinct from its low-$j$ counterpart. This difference - and not the pinning force - then leads to nonlinear, or even hysteretic, $V(I)$ in measurements over a wide range of $I$.\cite{lo,kunchur,weffi,dbbook}

The SNEQ mixed state has different backgrounds at low $T$ and at high $T$. At low $T$, as modeled by Kunchur,\cite{kunchur} the electron-phonon collisions are too infrequent to prevent electron heating (EH) to a temperature $T^*$ above the phonon temperature $T_0$, which leads to a thermal
quasiparticle distribution function that is set by $T^*$ rather than $T_0$. This causes an expansion of vortex cores. Close to  $T_c$, the dominant effect is the time variation of the superconducting order parameter $\Delta$ while the heating is negligible, and the distribution function acquires a nonthermal form as calculated by Larkin and Ovchinnikov (LO).\cite{lo} In consequence, vortex cores shrink. A detailed consideration of $V(I)$ in the two regimes\cite{weffi,dbbook} supported that: EH was identified at low $T$, and the
LO effect close to $T_c$. However, this conclusion relied on a somewhat intricate numerical analysis, which called for a more obvious proof of viability in order to rule
out other possible scenarios.\cite{henderxiao}

Recently, an alternative experiment provided a stronger support to the picture outlined above. This evidence came from dc measurements of the TFTE - the latter was introduced by Grigorieva {\it et al.} in Ref.~\onlinecite{grigorieva} - in a sample of $a$-NbGe.\cite{ottodiss,ottoprl}
The TFTE is a nonlocal phenomenon where the voltage response $V_\mathrm{nl}$, representative of vortex velocity, to a local $I$ in a mesoscopic film is measured in a remote region where $I=0$. In the TFTE, the flux coupling is transversal to the magnetic induction ${\bf B}$ (perpendicular to the film plane) and is caused by the in-plane repulsive intervortex interaction, which complements the longitudinal flux transformer effect of Giaver\cite{giaver} where the flux is coupled  along ${\bf B}$ over an insulating layer. First reports on the TFTE referred to low $I$ both in low-frequency ac (Ref.~\onlinecite{grigorieva}) and dc (Ref.~\onlinecite{helzel}) measurements, where it was found that $V_\mathrm{nl}$ was odd in $I$, i.e., $V_\mathrm{nl}(-I) = - V_\mathrm{nl}(I)$. This was a consequence of the local driving force $f_\mathrm{L} \propto I$ acting as a pushing or pulling locomotive for a train of vortices in the region of $I=0$.

In Ref.~\onlinecite{ottoprl}, this behavior - found again at low $I$ - changed dramatically at high $I$, where $V_\mathrm{nl}$ reversed sign to eventually become symmetric, exhibiting $V_\mathrm{nl}(-I) = V_\mathrm{nl}(I)$.
Remarkably, the sign of this even $V_\mathrm{nl}(I)$ was opposite at low $T$ and close to $T_c$. This
implied that the local SNEQ mixed states were completely different, which turned out
to be consistent with EH ($T \ll T_c$) and the LO effect ($T \lesssim T_c$) in the $I \neq 0$ region. Hence, the TFTE has offered a new possibility for distinguishing between EH and the LO effect in a manner that is free of numerical ambiguities mentioned before, since only the sign of $V_\mathrm{nl}$ has to be measured. The cause of $V_\mathrm{nl}$ with EH or the LO effect in the $I \neq 0$ region can be described by generalizing the magnetic-pressure model of Ref.~\onlinecite{helzel} to $f_\mathrm{dr}$ which is different from $f_\mathrm{L}$ and depends on the type of the local SNEQ.\cite{ottoprl} At low $T$, the origin of $f_\mathrm{dr}$ is a $T$ gradient at the interface of the $I \neq 0$ and $I=0$ regions, so $V_\mathrm{nl}$ is the consequence of a Nernst-like effect.\cite{huebenerbook} Close to $T_c$, vortices are driven by a Lorentz-like force induced at the interface and stemming from a novel enhancement of diamagnetism in the LO state relative to that in equilibrium.

In this paper, we give a timely account of other results of the experiment of Ref.~\onlinecite{ottoprl}.
These refer to eight temperatures from $0.75 \, \mathrm{K} \leq T \leq 2.80 \, \mathrm{K}$ (i.e., $0.26 \leq t \leq 0.95$, where $t=T/T_c$) and the whole range of applied magnetic field $B_\mathrm{ext}$ where the TFTE could be observed at a given $T$.\cite{ottodiss} EH persists up to 2 K ($t=0.68$) above which the LO effect takes place. The $T$ evolution of the SNEQ vortex dynamics is presented through changes in a characteristic high-$I$ voltage $V_\mathrm{nl}^*$. In order to account for the phenomenon quantitatively, $V_\mathrm{nl}^*$ is combined with the nonlocal resistance $R_\mathrm{nl}=V_\mathrm{nl}/I$ which is defined for the low-$I$ linear response regime and contains information on the pinning efficiency. Quantities characteristic of the TFTE with a given local SNEQ are traced in $T$ ranges of their relevance. These are the vortex transport entropy $S_\phi$ below 2 K, and the nonequilibrium magnetization ($M$) enhancement $\delta M$ in the LO state above 2 K. A special attention is paid to results at 2 K, where the LO effect is modified by EH above a certain $I$, which leaves a clear signature only in $V_\mathrm{nl}(I)$.

\section{Experiment}
\label{experiment}

\begin{figure}
\includegraphics[width=65mm]{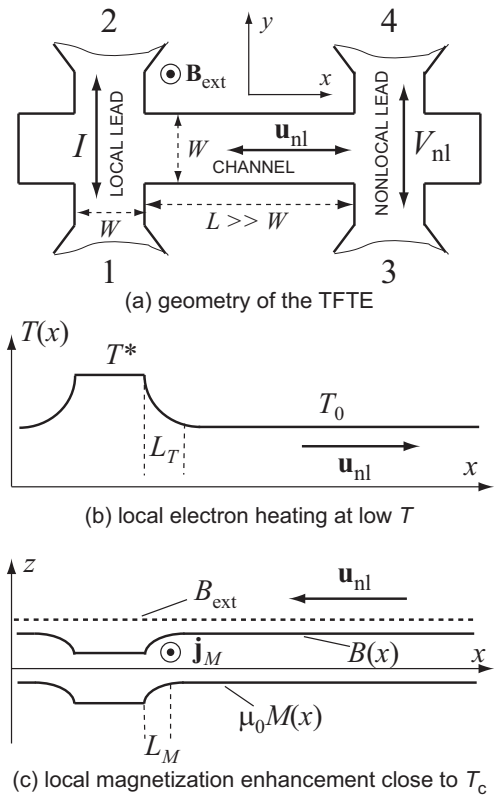}
\caption{(a) Schematic representation (not to scale) of the TFTE geometry, as used in 
Ref.~\onlinecite{ottoprl}. $B_\mathrm{ext}$ is applied perpendicularly to the ($x$-$y$) film plane. $I$ is passed between the contacts 1 and 2, and $V_\mathrm{nl}$ is measured between the contacts 3 and 4. (b) Temperature profile along the sample in the regime of EH in the local lead. (c) Profiles of $B_\mathrm{ext}$, $B$ and $\mu_0 M$ (all in the $z$ direction) along the sample, and consequent ${\bf j}_{M}$, in the regime of the LO effect in the local lead. In (b) and (c), the direction of ${\bf u}_\mathrm{nl}$ does not depend on the polarity of $I$.}
\label{sample}
\end{figure}

The sample of Ref.~\onlinecite{ottoprl} - a nanostructured $a$-Nb$_{0.7}$Ge$_{0.3}$ thin film - was produced by combining electron-beam lithography and magnetron sputtering onto an oxydized
Si substrate.\cite{ottodiss} The layout of the sample is presented schematically in Fig.~\ref{sample}(a). The film thickness is $d=40$ nm, the width is $W=250$ nm (in and around the channel) and the channel length is $L=2$ $\mu$m. The relevant coordinate system (with unit vectors $\hat{{\bf x}}$, $\hat{{\bf y}}$, $\hat{{\bf z}}$) is indicated. ${\bf B}_\mathrm{ext}= B_\mathrm{ext} \hat{{\bf z}}$ is perpendicular to the film plane. In measurements of
$V_\mathrm{nl}(I)$, one applies $\pm |I|$ between the contacts 1 and 2 (local lead).
The corresponding $|\bf{j}|$ decays exponentially away from the local lead, over a characteristic length $~\sim W/\pi \ll L$.\cite{grigorieva,ottodiss} Vortices in the channel are pressurized by the locally driven ones,\cite{helzel,ottoprl} and move along the channel at nonlocal velocity
${\bf u}_\mathrm{nl} = \pm |u_\mathrm{nl}| \hat{{\bf x}}$. This induces an electric field
${\bf E} = {\bf B} \times {\bf u}_\mathrm{nl}$ that is measured as $\pm V_\mathrm{nl}$ between the contacts 3 and 4 (nonlocal lead). The direction of ${\bf u}_\mathrm{nl}$, and consequently the sign of $V_\mathrm{nl}$, depends on the type of $f_\mathrm{dr}$, which will be addressed in Section~\ref{tfte}.

The same sample is used to measure the local dissipation. In this case,
$I$ is passed between 1 and 3, and the local voltage drop $V_\mathrm{l}$ is measured between 2 and 4.
Since $W$ is the sample width for all current paths (apart from a weak modulation of ${\bf j}$ in the local-lead area adjacent to the channel), $j \approx I / Wd$ is effectively the same both for measurements of $V_\mathrm{nl}$ and $V_\mathrm{l}$, which permits to use $V_\mathrm{l}(I)$ as a representative of the local vortex dynamics for $V_\mathrm{nl}(I)$ at the same $T$ and $B_\mathrm{ext}$.
Measurements of $V_\mathrm{l}$ also provide important parameters of the sample,\cite{kestsuei}
which are: $T_c=2.94$ K, the normal-state resistivity
$\rho_\mathrm{n}=1.82$ $\mu\Omega$m, the diffusion constant $D=4.8 \times 10^{-5}$ m$^2$/s, $-(dB_{c2}/dT)_{T=T_c}=2.3$ T/K, where $B_{c2}$ is the equilibrium upper critical magnetic field, and the Ginzburg-Landau parameters $\kappa=72$, $\xi(0)=7.0$ nm, and $\lambda(0)=825$ nm.\cite{ottoprl,ottodiss}
The low pinning, characteristic of  $a$-NbGe, allowed for dc measurements of $V_\mathrm{nl} \sim 10 - 200$ nV, which was at the level of $R_\mathrm{nl} \sim 0.1$ $\Omega$ in the low-$I$ linear regime. All measurements were carried out in a standard $^3$He cryostat.

\section{Local and nonlocal dissipation vs nonequilibrium vortex dynamics}
\label{neqstate}

In this Section, we give a brief overview of the SNEQ vortex-motion phenomena in $a$-NbGe films. Due to the simplicity of vortex matter and weak pinning in these systems,\cite{dbbook}  the discussed topics are related to fundamental issues of vortex dynamics rather than to sample-dependent pinning or peculiar vortex structure in exotic superconductors. We discuss limitations in the reliability of information that can be extracted from $V_\mathrm{l}(I)$ only, and the potential of $V_\mathrm{nl}(I)$ in identifying the microscopic processes behind an SNEQ mixed state.

\subsection{Types of SNEQ in vortex motion}
\label{types}

\begin{figure}
\includegraphics[width=75mm]{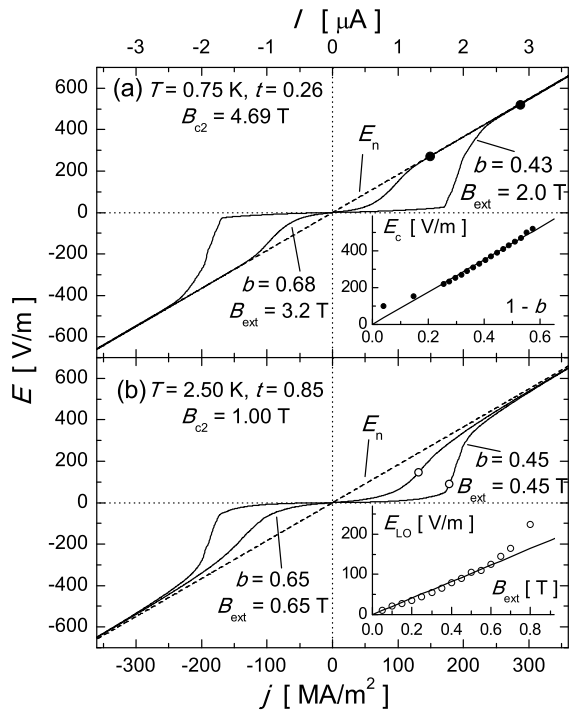}
\caption{Local $E(j)$ curves (solid lines), with $E_\mathrm{n} =\rho_\mathrm{n} j$ shown by the dashed lines. The values of $B_\mathrm{ext}$ and $b$ are given in the legends. (a) $T=0.75$ K, where $B_{c2}=4.69$ T. The solid circles represent $E_c$ which is in the inset plotted against ($1-b$) together with a linear fit (solid line) given by $E_\mathrm{c0}=900$ V/m. (b) $T=2.5$ K, where $B_{c2}=1 $ T. The open circles display $E_\mathrm{LO}$, in the inset plotted against $B_\mathrm{ext}$ together with a linear fit (solid line) corresponding to
$u_\mathrm{LO}=205$ m/s.}
\label{localonly}
\end{figure}

In Fig.~\ref{localonly}, we plot exemplary (nonhysteretic) $E(j)$ curves extracted from $V_\mathrm{l}(I)$ of the sample under discussion. The corresponding $I$ is shown on the top axis, the simple conversion being
$I \, [\mu \mathrm{A}] \leftrightarrow   j \, [100 \, \mathrm{MA/m}^2]$.
We choose two characteristic temperatures where the SNEQ is well defined, these are: (a) for EH, $T=0.75$ K ($t=0.26$, $B_{c2}=4.69$ T), and (b) for the LO effect, $T=2.5$ K ($t=0.85$, $B_{c2}=1$ T). The values of $b=B_\mathrm{ext}/B_{c2}$ are selected to demonstrate the cases of relatively strong ($b \sim 0.43-0.45$) and weak ($b \sim 0.65-0.68$) nonlinearities in $V_\mathrm{l}(I)$ at both temperatures.

At first sight, there is no obvious difference between the curves in Figs.~\ref{localonly}(a) and \ref{localonly}(b), but a closer look reveals that those in Fig.~\ref{localonly}(a) exhibit slightly sharper changes of curvature than their high-$T$ counterparts. A difference can also be noted at high dissipation where $E \lesssim E_\mathrm{n} = \rho_\mathrm{n} j$. In Fig.~\ref{localonly}(a), there is an electric field $E_c$, appearing at moderate $j$ and indicated by the solid circles, above which $E = E_\mathrm{n}$ within $~0.1$ \%. In contrast, the curves in Fig.~\ref{localonly}(b) slowly creep toward $E_\mathrm{n}$ but stay below by more than 1 \% over the whole range of $j$.  Thus, there are some features which point to different origins of the two types of $E(j)$, but these are barely visible and therefore difficult to spot.

Another way of determining the physics behind such a nonlinear $E(j)$ is to analyze the set of curves at a same $T$ numerically.\cite{kunchur,weffi,dbbook,ottodiss} At low $T$, one can concentrate on steep jumps of $E(j)$ at low $b$ by the method of Ref.~\onlinecite{kunchur}, or can address the high-$E$ part in the spirit of Ref.~\onlinecite{weffi}
for all $b$, both approaches being based on the assumption of a change $B_{c2}(T_0) \rightarrow B_{c2}(T^*)$ due to EH. The latter method results in a determination of $E_c$ which, according to a model based on the $b$ dependence of the Gibbs free energy density close to $B_{c2}$,\cite{weffi,dbbook} should be well approximated by $E_c = E_{c0}(1-b)$. The result of this procedure for $V_\mathrm{l}(I)$ at $T=0.75$ K is shown in the inset to Fig.~\ref{localonly}(a).\cite{ottodiss} The extracted $E_c$ is displayed by the solid circles, and the solid line is a linear fit with $E_{c0}=900$ V/m.\cite{errbars} This analysis also clarifies the meaning of $E_c$: at $E=E_c$, the heating destroys superconductivity, i.e., $T^* = T_c(B_\mathrm{ext})$, or, equivalently, 
$B_\mathrm{ext} = B_{c2}(T^*)$.

The framework for analyzing $E(j)$ close to $T_c$ is
different.\cite{weffi,dbbook,ottodiss} In this case, one uses the LO expression for $E(j)$,  which describes a  dynamic reduction of the vortex-motion viscosity coefficient $\eta$.\cite{lo} The main quantity to be determined from $E(j)$ is the characteristic
LO electric field $E_\mathrm{LO} = u_\mathrm{LO}B$, where $u_\mathrm{LO}$ is the LO vortex velocity.
The positions of $E_\mathrm{LO}$ are in Fig.~\ref{localonly}(b) shown by the open circles, and the same symbols are used for plotting $E_\mathrm{LO}$  against $B_\mathrm{ext}$  in the corresponding inset. The approximation
$B \approx B_\mathrm{ext}$ is justified by $|M| \ll B_\mathrm{ext}$ for a high-$\kappa$ superconductor in the mixed state. The solid line is a linear fit with $u_\mathrm{LO} = 205$ m/s.

The extracted $E_c$ and $E_\mathrm{LO}$ follow the predicted dependences reasonably well but still not as good as in Ref.~\onlinecite{weffi} - where measurements were carried out on a 5 $\mu$m wide microbridge - which also holds for the overall agreement of the shape of the experimental $E(j)$ with the models outlined above.\cite{ottodiss} We believe that the main reason for this discrepancy lies in the characteristic times involved in establishing an SNEQ in such narrow strips. This can be demonstrated by the following consideration. 
The time required for a creation/destruction of the LO state is the relaxation time of nonequilibrium quasiparticle excitations, which is close to $T_c$ given by 
$\tau_\varepsilon \sim \tau_\mathrm{e,ph} k_\mathrm{B} T_c / | \Delta |$ with $\tau_\mathrm{e,ph}$ being the electron-phonon scattering time and $k_\mathrm{B}$ the Boltzmann constant.\cite{schmid}
For the given $u_\mathrm{LO} \approx 205$ m/s and other sample parameters, $\tau_\varepsilon$ is around 
1.5 ns.\cite{ottodiss} On the other hand, the time of vortex traversal across our sample in the LO regime is of the order of $\tau_W \sim W /u_\mathrm{LO} \approx 1.2$ ns, i.e., about the same as $\tau_\varepsilon$. This was not the case in Ref.~\onlinecite{weffi} where the LO state fully developed because of $\tau_\varepsilon \ll \tau_W$. A similar analysis, leading to the same conclusions, can be done for EH as well.

There are several messages of the above overview. First, the shape of $E(j)$ can be almost the same for distinct SNEQ mixed states, with hardly detectable differences. Second, numerical analyses can also be of limited reliability if the samples are very small. Moreover, any combination of these qualitative and quantitative approaches could fail to give a proper answer on the nature of an SNEQ when $T$ is neither low nor close to $T_c$, i.e., when a competition between EH and the LO effect may occur. The latter point will be addressed more closely in Section~\ref{intert}.

\subsection{TFTE vs local vortex dynamics}
\label{tfte}

Local $V(I)$ curves in the mixed state are generally monotonic and odd in $I$, apart from their possible weakly hysteretic behavior at low $b$.\cite{dbbook}  In contrast, $V_\mathrm{nl}(I)$ measured over a wide range of $I$ is nonmonotonic and at first glance lacks any even or odd symmetry.\cite{ottoprl,ottodiss} This is a consequence of  different contributions to $f_\mathrm{dr}$, which do not have the same $I$ dependence. At low $j$, the driving force $f_\mathrm{dr} = f_\mathrm{L}$ is purely electromagnetic, as the mixed-state thermodynamics in the local lead remains essentially intact. For that reason, $f_\mathrm{L}$ is odd in $j$, and the resulting $V_\mathrm{nl}(I)$ is odd too. On the other hand, SNEQ at high $j$ in the local lead is a thermodynamic state different from that in the channel, and it is this difference which produces the SNEQ part of $f_\mathrm{dr}$. This part does not depend on the sign of $j$ because the creation of a local SNEQ is set by $|j|$, and the resulting $V_\mathrm{nl}(I)$ cannot be odd. Consequently, a wide-range sweep from $-I$ to $+I$ results in $V_\mathrm{nl}(I)$ of a rich structure,\cite{ottoprl,ottodiss} which is advantageous in determining the physics behind an SNEQ mixed state.

A generalization of the model of Ref.~\onlinecite{helzel} for $V_\mathrm{nl}$ as a response to $f_\mathrm{dr}$ can reasonably well account for the complexity of $V_\mathrm{nl}(I)$ in Ref.~\onlinecite{ottoprl}. This approach relies on a plausible assumption that vortices in the local lead push or pull those in the channel due to intervortex repulsion, and that the vortex matter is incompressible against this uniaxial magnetic pressure. The pressurizing occurs at the $W$-wide interface of the local lead and the channel, see Fig.\ref{sample}(a). The pushing/pulling force is produced by $n_\phi WX$ vortices under the direct influence of $f_\mathrm{dr}$, where $X$ is the distance over which $f_\mathrm{dr}$ extends in the $x$ direction, and $n_\phi = B / \phi_0$ is the vortex density. The number of vortices in the channel is $n_\phi WL$, and the motion of each of these vortices is damped by a viscous drag (per unit vortex length) $\eta u_\mathrm{nl}$. The driving and damping forces are balanced, i.e., $f_\mathrm{dr} \times (n_\phi WX) = (\eta u_\mathrm{nl}) \times (n_\phi WL)$, hence $u_\mathrm{nl} = f_\mathrm{dr} X / \eta L$ determines $V_\mathrm{nl} = B u_\mathrm{nl} W$. As before, we can approximate $B \approx B_\mathrm{ext}$ for a high-$\kappa$ superconductor to obtain the nonlocal current-voltage characteristics

\begin{equation}
\label{modeleq}
V_\mathrm{nl}(I) = \frac{W B_\mathrm{ext} X}{\eta L} f_\mathrm{dr}(I) \; .
\end{equation}
This expression does not apply below a certain magnetic field $B_d(T)$ that originates in the pinning
in the channel, and also  in the vicinity of the phase transition at $B_{c2}(T)$. 
More precisely, $V_\mathrm{nl}=0$ below $B_d$ and close to $B_{c2}$, so the TFTE is always restricted to a range of $B_\mathrm{ext}$.\cite{grigorieva,ottoprl,helzel,ottodiss}

When ${\bf f}_\mathrm{dr}= \pm |f_\mathrm{L}| \hat{{\bf x}} = \pm (\phi_0 |I|/  Wd) {\bf \hat{x}}$ for the sample orientation in Fig.\ref{sample}(a), vortices in the local lead contribute to $f_\mathrm{dr}$ over the whole width, and $X=W$.
This results in

\begin{equation}
\label{vlorentz}
V_\mathrm{nl} (I)  = \frac{W B_\mathrm{ext} \phi_0}{\eta L d} I = R_\mathrm{nl} I \;.
\end{equation}
The above expression satisfies $V_\mathrm{nl}(-I)= - V_\mathrm{nl}(I)$ and as well introduces $R_\mathrm{nl}$ as a measure of the TFTE efficiency. $R_\mathrm{nl}$ depends entirely
on the channel properties, in particular on $\eta$ for vortices out of SNEQ. In Ref.~\onlinecite{helzel},
the use of a theoretical $\eta = \eta_f$ of pining-free flux flow reproduced the experimental values of $R_\mathrm{nl}$ when the pinning was negligible  (close to $T_c$). 
When the pinning became stronger, at low temperatures, $R_\mathrm{nl}$ was lower than that calculated for pure flux flow but remained constant, i.e., $V_\mathrm{nl}(I)$ was still linear. This property was assigned to the motion of a depinned fraction of vortices in the channel, which was affected by a shear with the pinned (or slower) vortices but responded linearly to $I$.\cite{helzel} These effects can be parametrized by introducing an effective $\tilde{\eta} > \eta_f$ which does not depend on $I$.

We now turn to the TFTE at low $T$, where EH underlies the local SNEQ. The corresponding $T(x)$ is sketched in Fig.~\ref{sample}(b). In the local wire, $T=T^*$ which over a length
$L_T$ drops to $T=T_0$ in the channel. The driving force is a thermal force produced by the $T$ gradient,\cite{huebenerbook} and this behavior belongs to the class of Nernst-like effects. More precisely, ${\bf f}_\mathrm{dr} = {\bf f}_{T} =- S_\phi (\partial T / \partial x) \hat{{\bf x}} \approx
S_\phi [(T^* - T_0) / L_T] \hat{{\bf x}} $ is always in the positive $x$ direction because
$S_\phi > 0$, i.e., it drives vortices away from the local lead. With $X \approx L_T$, one obtains
\begin{equation}
\label{vheating}
V_\mathrm{nl} (I) = \frac{S_\phi  R_\mathrm{nl} d}{\phi_0} \delta T (I) \;  ,
\end{equation}
where $\delta T (I)= T^*(I) - T_0$ and $R_\mathrm{nl}$ is the same as in Eq.~(\ref{vlorentz}). Here, $V_\mathrm{nl}(-I)= V_\mathrm{nl}(I)$ because $f_T$ stems from the difference of thermodynamic potentials in the local and nonlocal regions. Notably, $L_T$ does not appear in Eq.~(\ref{vheating}) but it is still an important parameter in context of the magnitude of $f_{T}$ and the applicability of the model - which requires $L_T \ll L$.
For the sample of Ref.~\onlinecite{ottoprl}, this condition is fulfilled because the estimated $L_T$ 
in the relevant $T$ range of measurements (0.75 - 1.5 K) is between 125 nm (at 1.5 K) and 295 nm (at 0.75 K).\cite{ottodiss}

As explained before, the SNEQ close to $T_c$ corresponds to the LO effect. It follows from a calculation in Ref.~\onlinecite{ottoprl}, which is presented in more detail in Appendix \ref{LOmag}, that the nonequilibrium diamagnetic $|M|=|M_\mathrm{neq}|$ in the LO state is larger than $|M|=|M_\mathrm{eq}|$ in equilibrium. This results in spatially nonuniform profiles of $\mu_0 M$ and $B$, where $\mu_0 = 4 \pi \times 10^{-7}$ Vs/Am, as depicted in Fig.~\ref{sample}(c). The nonuniformity of $M$ creates a current density ${\bf j}_M = (\nabla \times {\bf M})_y \hat{{\bf y}}$ at the interface that stretches over $X=L_M$. Therefore, $j_M = - (\partial M / \partial x) \approx (M_\mathrm{neq} - M_\mathrm{eq}) / L_M  < 0$, i.e., ${\bf j}_M$ is always in the negative $y$ direction. This leads to a Lorentz-like force ${\bf f}_\mathrm{dr} = {\bf f}_M = -|j_M| \phi_0 \hat{{\bf x}}$ that drives vortices toward the local lead. Hence,
\begin{equation}
\label{vlo}
V_\mathrm{nl} (I) = [R_\mathrm{nl} d] \delta M(I) \;  ,
\end{equation}
where $\delta M (I)= | M_\mathrm{neq}(I) - M_\mathrm{eq} |$ and $R_\mathrm{nl}$ is again the same is in Eq.~(\ref{vlorentz}). Since $M$ also determines thermodynamic potentials,  $V_\mathrm{nl}(-I)= V_\mathrm{nl}(I)$ but of the sign which is opposite to that in Eq.(\ref{vheating}). As before, $L_M$ drops out from the expression for $V_\mathrm{nl}(I)$ but should be addressed because it is an important parameter in both the magnitude and the extent of $f_M$. The issue of $L_M$ is, however, less straightforward than that of $L_T$.

In Ref.~\onlinecite{ottoprl}, it was shown that the reason for $\delta M$ was a nonequilibrium gap enhancement near the vortex cores in the LO state. The net effect is
an increase of the magnetic moment of a single-vortex Wigner-Seitz cell. In the equatorial plane, the dipole magnetic field of an individual cell opposes $B_\mathrm{ext}$ in other cells and in this way reduces $B$.
Therefore, the larger the gap enhancement, the larger the diamagnetic response.
The gap enhancement occurs at the expense of quasiparticles within the cores, which have energies below
the maximum $| \Delta |_\mathrm{max}$ of $| \Delta |$ in the intervortex space. These quasiparticles can penetrate into the surrounding superfluid by Andreev reflection only, i.e., up to a distance of about the coherence length $\xi$ - which is the first candidate for $L_M$. On the other hand, this process is a single-vortex property, whereas $\delta M$ requires a many-vortex system.  The second candidate is 
$L_\varepsilon = \sqrt{ D \tau_\varepsilon}$ but this length is more specific of quasiparticles with energies above
$| \Delta |_\mathrm{max}$. There is, however, a third candidate as well. This is the intervortex distance $a_0 \sim (\phi_0 /B)^{1/2}$ which plays a crucial role in the screening of $B_\mathrm{ext}$ as explained above. Thus, we believe that the proper estimate for $L_M$ is $a_0$, although this matter is certainly still open to debate. In any case, $L_M \ll L$ holds.

\section{Results and discussion}
\label{resdis}

\begin{figure}
\includegraphics[width=75mm]{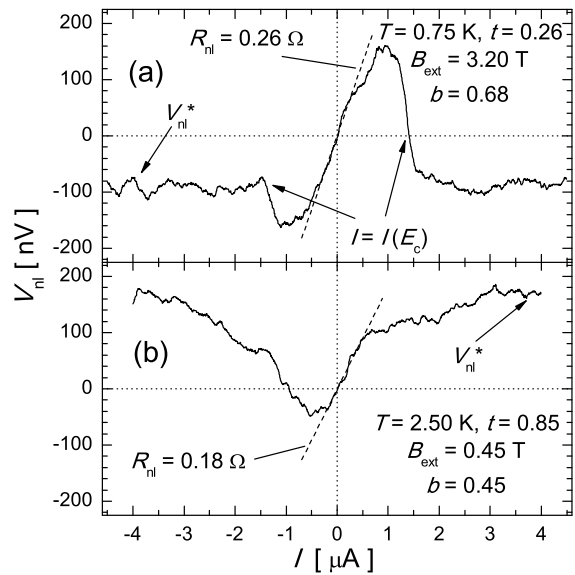}
\caption{$V_\mathrm{nl}(I)$ in the presence of (a) EH, and (b) the LO effect in the local lead, for measurements where the overall TFTE strength is maximal in the two regimes. Slopes of the linear dashed lines determine $R_\mathrm{nl}$. The arrows point to $V_\mathrm{nl}^*$ and, in (a), also to $V_\mathrm{nl}$ at $I(E_c)$, see the text. The values of important parameters are given in the legends, and the corresponding local dissipation is presented in Fig.~\ref{localonly}.}
\label{rawehlo}
\end{figure}

Henceforth, we turn to experimental results which support the concepts presented above. General trends in $V_\mathrm{nl}(I)$ are demonstrated using experimental curves at $(T,B_\mathrm{ext})$ points where the
TFTE is maximal for the two local SNEQ regimes. These are shown in Fig.~\ref{rawehlo}: (a) for EH, at $T=0.75$ K and $B_\mathrm{ext}=3.2$ T, and (b) for the LO effect, at $T=2.5$ K and $B_\mathrm{ext}=0.45$ T.
The $B_{c2}$ values are 4.69 T and 1 T, respectively, thus $t=0.26$ and $b=0.68$ in (a), and
$t=0.85$ and $b=0.45$ in (b). Note that the corresponding local dissipation curves are displayed in Fig.~\ref{localonly}.

We first return to Fig.~\ref{sample}(a) to explain the signs in $V_\mathrm{nl}(I)$ plots.
${\bf B}_\mathrm{ext}$ is always directed as shown, $I > 0$ represents ${\bf j}$ downwards, and
$V_\mathrm{nl} > 0$ means ${\bf u}_\mathrm{nl}$ leftwards, i.e., towards the local lead. 
The $V_\mathrm{nl}(I)$ saturates at high $I$ both in Fig.~\ref{rawehlo}(a) and Fig.~\ref{rawehlo}(b), but the sign of the saturation voltage is opposite in the two regimes. 
The saturation occurs for most of measured $V_\mathrm{nl}(I)$, except when there is a physical reason (see Section~\ref{intert}) for the saturation to be shifted beyond the maximum used $I$ of $4 - 5$ $\mu$A.
Without introducing a significant error, instead of characterizing $V_\mathrm{nl}$ strictly by the saturation value,  we use $V_\mathrm{nl}^*= V_\mathrm{nl}(|I|=4 \,\mu \mathrm{A})$, indicated by the arrows, to represent the strength of the TFTE at a local SNEQ. Another measure of the (overall) TFTE efficiency is $R_\mathrm{nl}$ which can be extracted from the antisymmetric part $V_\mathrm{nl}=R_\mathrm{nl}I$ corresponding to $f_\mathrm{dr} = f_\mathrm{L}$ at low $I$, as indicated by the dashed lines.

The difference between the curves in Figs.~\ref{rawehlo}(a) and \ref{rawehlo}(b) becomes striking at high $I$, in contrast to that between the curves in Figs.~\ref{localonly}(a) and \ref{localonly}(b). This implies  availability of information from $V_\mathrm{nl}(I)$ without any in-depth analysis. For example, at $|I| \approx 1.5$ $\mu$A, where $E = E_c$ in Fig.~\ref{localonly}(a), $V_\mathrm{nl}(I)$ in Fig.~\ref{rawehlo}(a) either changes sign 
(for $I > 0$) or starts to be flat when  $I <0$ strengthens further. The asymmetry originates in ${\bf f}_T$ and ${\bf f}_L$ acting in the same direction for $I<0$, and in the opposite directions when $I>0$.
The same $V_\mathrm{nl}^*$ for $I <0$ and $I > 0$ is a consequence of $f_\mathrm{L}=0$ for $E > E_c$.
Besides being completely different, the $V_\mathrm{nl}(I)$ in Fig.~\ref{rawehlo}(b) exhibits no sharp features. This is consistent with the LO effect not leading to a destruction of superconductivity in the range
of $I$ used, as already pointed out in Section~\ref{neqstate}.

\begin{figure}
\includegraphics[width=75mm]{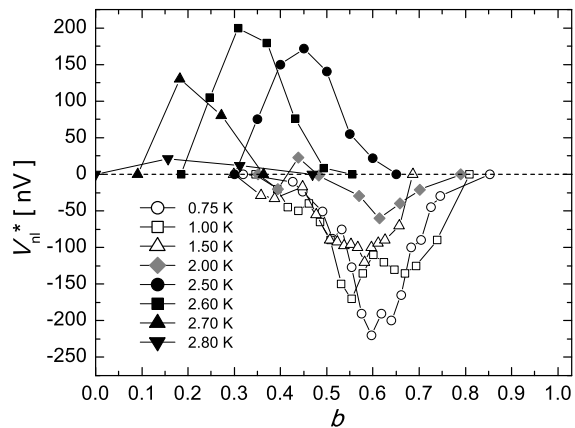}
\caption{Plot of $V_\mathrm{nl}^*$ vs $b$ for all $T$ where TFTE data were collected, as indicated in the legend. For the local EH (open symbols), $V_\mathrm{nl}^* <0$, and for the local LO effect (solid symbols), $V_\mathrm{nl}^* > 0$. At $T=2$ K (grey diamonds), there is no proper saturation of $V_\mathrm{nl}(I)$, and $V_\mathrm{nl}^*$ does not exhibit a well-defined behavior.}
\label{vstar}
\end{figure}

We shall consider these and other issues in more detail later, but it is worthwhile to begin by  a simple plot of $V_\mathrm{nl}^*$ against $b$ for all $T$ where our TFTE data were
collected.\cite{ottodiss} This is done in Fig.~\ref{vstar}. It is seen that $V_\mathrm{nl}^* < 0$ for $T=0.75,1,1.5$ K, which implies the local EH, and $V_\mathrm{nl}^* > 0$ at $T=2.5,2.6,2.7,2.8$ K, suggesting the LO effect in the local lead. There is, however, an intermediate behavior at $T=2$ K, where $V_\mathrm{nl}(I)$ does not show a proper saturation and $V_\mathrm{nl}^*$ does not clearly belong to either of the two regimes. These three cases are  addressed separately below.

\subsection{TFTE well below $T_c$}
\label{lowt}

\begin{figure}
\includegraphics[width=75mm]{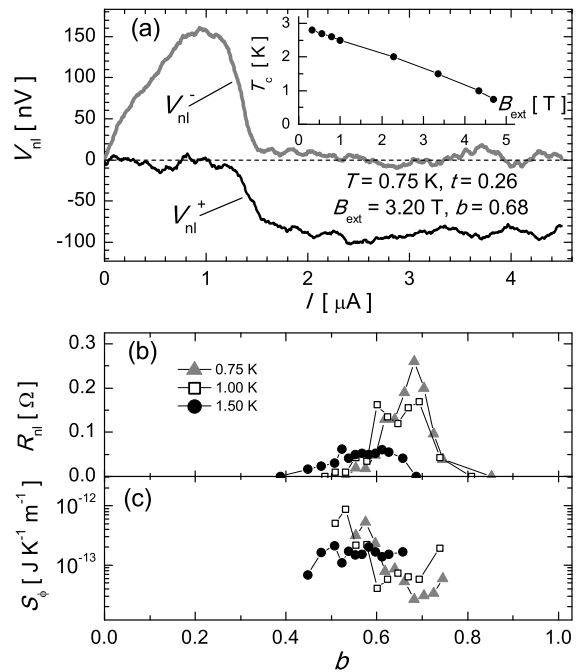}
\caption{(a) $V_\mathrm{nl}^{+}(I)$ and $V_\mathrm{nl}^{-}(I)$, as indicated, for the $V_\mathrm{nl}(I)$ in Fig.~\ref{rawehlo}(a). Inset: Experimental $T_c(B_\mathrm{ext})$. (b) $R_\mathrm{nl}$ vs $b$ for measurements where the
SNEQ in the local lead is caused by EH. (c) $S_\phi$ against $b$, plotted with the same symbols as in (b) and calculated as explained in the text. For (b) and (c), $T$ is indicated in the legend to (b).}
\label{lowtres}
\end{figure}

In order to understand different contributions to $V_\mathrm{nl}(I)$, it is appropriate do decompose it
into $V_\mathrm{nl}^{\pm} (I) = [V_\mathrm{nl}(I) \pm V_\mathrm{nl}(-I)]/2$. The symmetric part $V_\mathrm{nl}^{+}$ is  representative of the thermodynamic forces $f_T$ and $f_M$, whereas the antisymmetric part $V_\mathrm{nl}^{-}$ accounts for the electromagnetic force $f_\mathrm{L}$. The result of this approach for the $V_\mathrm{nl}(I)$ in Fig.~\ref{rawehlo}(a) is displayed in Fig.~\ref{lowtres}(a),
and is typical of the low-$T$ regime. $V_\mathrm{nl}^{-} \propto I$ is found at low $I$, with
$V_\mathrm{nl}^{+}$ at the same time being very small, and this suggests $f_\mathrm{dr} \approx f_\mathrm{L}$.
As $I$ increases, $V_\mathrm{nl}^{-}$ at some point starts to decrease and 
$V_\mathrm{nl}^{+} <0$ simultaneously to grow, which implies a transition towards $f_\mathrm{dr} \approx f_T$.
Eventually, around $I(E_c) \approx 1.5$ $\mu$A, $V_\mathrm{nl}^{-}$ drops to zero and $V_\mathrm{nl}^{+}$ approaches a constant value. 

The $T^*(I)$ characteristics exemplified in Ref.~\onlinecite{ottoprl}
indicates a one-to-one correspondence of EH in the local lead and $V_\mathrm{nl}^{\pm}(I)$. Analysis of
the $V_\mathrm{l}(I)$ in the superconducting state [$T^* < T_c(B_\mathrm{ext})$] by the method of
Ref.~\onlinecite{weffi} connotes that $T^*(I)$ first increases
slowly and then jumps very steeply in the $I$ window where the above-discussed steep changes of
$V_\mathrm{nl}^{\pm}(I)$ occur.\cite{ottodiss,ottoprl} The high-$I$ part, where $V_\mathrm{nl}^{-}=0$ and $V_\mathrm{nl}^{+} \approx \mathrm{const.}$, corresponds to the normal state in the local lead. Noise measurements\cite{ottodiss,henny} in this regime indicate a marginal increase of $T^*$ with increasing $I$, hence one can assume $T^* \approx T_c(B_\mathrm{ext})$ regardless of $I$. Therefore, there is a relatively abrupt transition from $f_\mathrm{dr} \approx f_\mathrm{L}$
to $f_\mathrm{dr} \approx f_T$ when $T^*$ is close to $T_c(B_\mathrm{ext})$ [the experimental results for which
are shown in the inset to Fig.~\ref{lowtres}(a)].

The strongest effect of $f_T$ occurs at $T_c(B_\mathrm{ext}) \lesssim T^*$, i.e., when the local lead is in the normal state. In this regime, vortices nucleate somewhere within the length $L_T$ away from the local lead,
move toward the channel due to the $T$ gradient, and push vortices in the channel. This situation is
different from that in conventional measurements of the Nernst effect,\cite{huebener,vidal} because here $T$ gradients are very strong ($\sim 1$ K/$\mu$m), the number of vortices under the direct influence of $f_T$ is small, and the voltage corresponds to the motion of vortices which are in an isothermal environment (the channel remains at $T=T_0$). Strong lateral temperature variations over $a$-NbGe microbridge films (also on oxidized Si) due to EH at low $T$ were also observed in a noise experiment.\cite{wenoise} This gives an additional support to the reality of spatially dependent separation of the electron temperature $T^*$ and the phonon temperature $T_0$ at least for the given substrate-film interface properties.\cite{bs}

In Fig.~\ref{lowtres}(b), we show $R_\mathrm{nl}(b)$ at $T=0.75,1,1.5$ K, i.e., for temperatures where the local SNEQ corresponds to EH  [the overall magnitude of $R_\mathrm{nl}(T)$ will be discussed later]. In Fig.~\ref{lowtres}(c), we use the same symbols to plot $S_\phi(b)$ obtained by inserting $R_\mathrm{nl}$, $V_\mathrm{nl}= | V_\mathrm{nl}^* |$  and $\delta T = [T_c(B_\mathrm{ext}) - T_0]$ into Eq.~(\ref{vheating}). The intricacy of the experimental situation has been outlined above, so it is not straightforward to analyze $S_\phi$ in terms of the Maki formula\cite{maki,kopnin} $S_\phi = \phi_0 |M_\mathrm{eq}|/T$ [where $M_\mathrm{eq} \approx (B_\mathrm{ext} - B_{c2})/2.32 \mu_0 \kappa^2$ for $B_\mathrm{ext}$ not much below $B_{c2}$] which applies to a weak $T$ gradient over the whole sample and no local destruction of superconductivity by heating. On the other hand, if $f_T$ is really the relevant $f_\mathrm{dr}$, then the extracted $S_\phi$ should still be reasonable in terms of the order of magnitude. This is indeed the case, since our $S_\phi$ does not depart significantly neither from the estimate by the Maki formula with $T = T_0$, giving
$S_\phi \sim 0.1-0.2 \times 10^{-12}$ Jm$^{-1}$K$^{-1}$, nor from the values in experiments  of Ref.~\onlinecite{huebener} (Nb films) and Ref.~\onlinecite{vidal} (Pb-In films), where it was found $S_\phi \sim 0.05-1.5 \times 10^{-12}$ Jm$^{-1}$K$^{-1}$ and $S_\phi \sim 0.2-5 \times 10^{-12}$ Jm$^{-1}$K$^{-1}$, respectively. Thus, we conclude that our results for the TFTE at low $T$ are consistent with the picture of local EH and the consequent Nernst-like effect.

\subsection{TFTE close to $T_c$}
\label{hight}

\begin{figure}
\includegraphics[width=75mm]{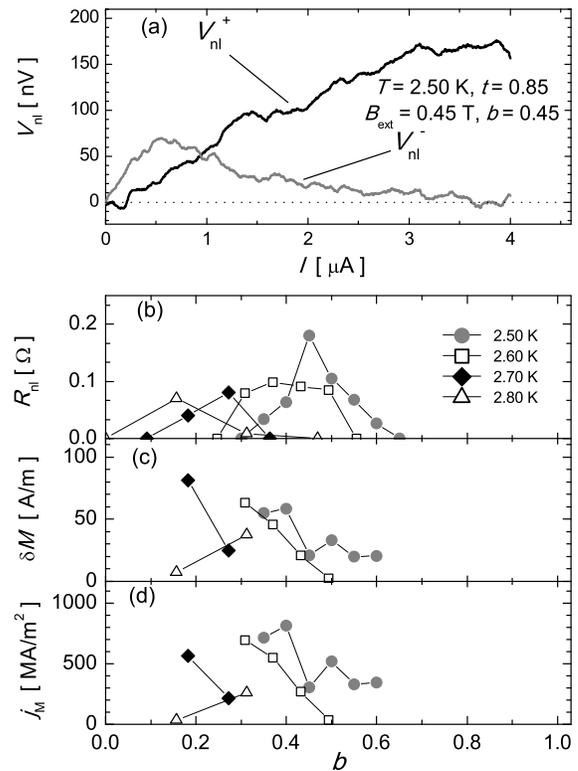}
\caption{(a) $V_\mathrm{nl}^{+}(I)$ and $V_\mathrm{nl}^{-}(I)$ for the $V_\mathrm{nl}(I)$ in Fig.~\ref{rawehlo}(b), as indicated. (b) $R_\mathrm{nl}$ vs $b$ for measurements where the SNEQ in the local lead is caused by the LO effect.
(c) Magnetization enhancement $\delta M(b)$, calculated using Eq.~(\ref{vlo}). (d) Interface current $j_M(b)$, extracted from $\delta M(b)$. For (b)-(d), $T$ is indicated in the legend to (b).}
\label{hightres}
\end{figure}

The method of analyzing  $V_\mathrm{nl}^{\pm} (I)$ can also be applied to the TFTE at $T \lesssim T_c$.
For the $V_\mathrm{nl}(I)$ in Fig.~\ref{rawehlo}(b), this results in $V_\mathrm{nl}^{+} (I)$ and
$V_\mathrm{nl}^{-} (I)$ displayed in Fig.~\ref{hightres}(a). Let us first discuss $V_\mathrm{nl}^{-} (I)$. As before,
$V_\mathrm{nl}^{-} \propto I$ at low $I$, but - in contrast to the low-$T$ behavior - this is followed by a slow decay of $V_\mathrm{nl}^{-}$ as $I$ increases, not by a sharp drop to zero. The linear part of $V_\mathrm{nl}^{-}(I)$ is again a consequence of $f_\mathrm{L}$ dominating in $f_\mathrm{dr}$ at low $I$, whereas the decrease of $V_\mathrm{nl}^{-}(I)$ at high $I$ can be explained by a reduction of $f_\mathrm{L}$ in the high-dissipation regime of vortex motion. Namely,
when $E \lesssim E_\mathrm{n}$, which can be a consequence either of an SNEQ or of $b \lesssim 1$ in a close-to-equilibrium situation, a significant fraction of $j$ is carried by quasiparticles.\cite{lo} This normal current does not lead to asymmetry in the profile of $\Delta$ around the vortex core, which is set by the supercurrent density $j_s$, and it therefore does not contribute to $f_\mathrm{L}$.\cite{superj} The observed progressive  reduction of $V_\mathrm{nl}^{-}(I)$ as $b$ grows\cite{ottoprl} is in support to this picture.

The main information about the SNEQ is contained in $V_\mathrm{nl}^{+}(I)$ which increases monotonically with increasing $I$ until it saturates. As explained before, $V_\mathrm{nl}^{+}$ represents $f_M$ that is given by $\delta M$ at $T=T_0$. As $I$ increases, $\delta M$ grows until the core shrinking reaches its limit\cite{lo} at
$\xi(t) (1-t)^{1/4}$, when the increase of $\delta M$ must saturate.\cite{ottoprl} This simple consideration explains the shape of $V_\mathrm{nl}^{+}(I)$ qualitatively. Quantitatively, we can use Eq.~(\ref{vlo}) and $R_\mathrm{nl}(b)$, shown in Fig.~\ref{hightres}(b), to calculate $\delta M(b)$. The result of this procedure is shown in Fig.~\ref{hightres}(c). It can be seen that $\delta M$ is around 50 A/m, which is a very small value corresponding to $\sim 60$ $\mu$T. However, $\delta M$ is not small on the scale of $|M_\mathrm{eq}|$ which is of the same order. Moreover, the gradient of $M$ occurs over a small distance of the intervortex spacing
$a_0 \sim (\phi_0 / B_\mathrm{ext})^{1/2}$ which - for the given $B_\mathrm{ext}$ range - takes values between 60 nm and 140 nm. The calculated interface current $j_M = \delta M / a_0$ is plotted against $b$ in Fig.~\ref{hightres}(d), where it can be seen that it is comparable to a typical $j$ in our experiment.

There are also other issues of relevance for the TFTE at $T \lesssim T_c$. In our measurements, SNEQ develops in the local-lead area adjacent to the channel, as well as in the $W$-wide parts of the local lead along the $y$ direction, see Fig.~\ref{sample}(a). The local lead widens up further away and $j$ is smaller there, which introduces additional interfaces of the SNEQ and close-to-equilibrium mixed states. In the presence of an SNEQ in the local lead, vortices do not simply traverse the SNEQ area (as they do when $f_\mathrm{dr} = f_\mathrm{L}$): they all move either away ($T \ll T_c$) or toward ($T \lesssim T_c$) it. This must modify vortex trajectories in order to maintain $n_\phi = B / \phi_0$ via complex vortex entry/exit paths in and around the SNEQ area. At $T \ll T_c$, the problem is less troublesome because the strongest effects occur when EH has destroyed superconductivity and there are no vortices in the SNEQ area. Close to $T_c$, on the other hand, there are vortices everywhere, their sizes and velocites being spatially dependent. Obviously, their trajectories must be such that a local growth of $n_\phi$ is prevented, as this would cost much energy due to the stiffness of a vortex system  against compression. Moreover, while there is experimental evidence for a triangular vortex lattice in the channel,\cite{welattice} this cannot be claimed for the SNEQ area where the above effects could cause a breakdown of the triangular symmetry. This may be complicated further by sample-dependent pinning landscape, edge roughness, etc., but our simple model can nonetheless still account for the main physics of the phenomenon. Another subject related to effect of the sample geometry on the magnitude of $\delta M$ is discussed in Appendix~\ref{qpdiff}.

Last but not the least, our results may have implications for other topics as well. We have shown that there are two thermodynamic forces that can incite vortex motion and set its direction. Gradients of $T$ and $M$ can be created and controlled by external heaters and magnets, and it therefore seems that a combination of these two approaches can be useful in elucidating the presence of vortices or vortexlike excitations in different situations. For instance, current debate on the origin of the Nernts effect in high-$T_c$ compounds\cite{koki} could benefit from supplements obtained in experiments based on applying a gradient of $M$ in an isothermal setup.

\subsection{TFTE at intermediate $T$}
\label{intert}

\begin{figure}
\includegraphics[width=75mm]{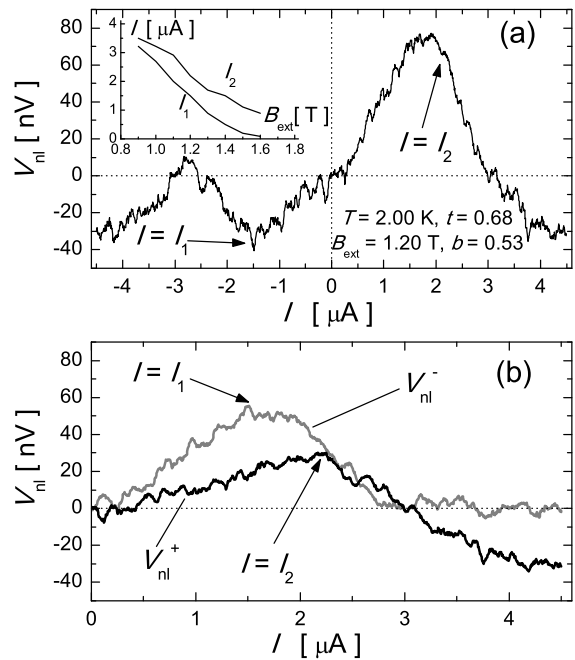}
\caption{(a) $V_\mathrm{nl}(I)$ at $T=2$ K ($t=0.68$, $B_\mathrm{ext}=1.2$ T, $b=0.53$) where neither EH nor the LO effect can give a conclusive description of $V_\mathrm{l}(I)$. (b) $V_\mathrm{nl}^{-} (I)$ and $V_\mathrm{nl}^{+} (I)$, as indicated. The latter exhibits a change of the sign, which is  suggestive of the appearance of EH on top of the LO effect which dominates at lower $I$. Inset to (a): $I_1$ and $I_2$, in the main panels indicated by the arrows, against $B_\mathrm{ext}$.}
\label{intertres}
\end{figure}

We have shown in previous Sections that the SNEQ mixed states at $T \ll T_c$ and $T \lesssim T_c$ have different physical backgrounds. However, the situation is less clear at intermediate $T$. For instance,
analysis of local $V(I)$ at $T=2$ K in Ref.~\onlinecite{weffi} was not conclusive, and these data were used only later in a qualitative consideration of another
phenomenon.\cite{weneqholes} The same applies to $V_\mathrm{l}(I)$ at $T=2$ K of this work, and this is
where the TFTE is crucial in determining the nature of the corresponding SNEQ mixed state.

In Fig.~\ref{intertres}(a), we present $V_\mathrm{nl}(I)$ at $T=2$ K ($t=0.68$) and $B_\mathrm{ext}=1.2$ T ($b=0.53$), the shape of which is markedly different from those in Fig.~\ref{rawehlo}. 
There are pronounced minima and maxima for both polarities of $I$, there are only indications of a saturation of $V_\mathrm{nl}(I)$ at the maximum current used, etc. A better understanding of the underlying physics can again be obtained from the corresponding $V_\mathrm{nl}^{-} (I)$ and $V_\mathrm{nl}^{+} (I)$, which are shown in  Fig.~\ref{intertres}(b). At $I < I_1$, there is a usual behavior
$V_\mathrm{nl}^{-} \propto I$, characteristic of the linear action of $f_\mathrm{L}$. Looking back at
Fig.~\ref{intertres}(a), one can see that $I=I_1$ corresponds to the minimum of $V_\mathrm{nl} (I)$ on the
$I<0$ side. $V_\mathrm{nl}^{+} (I)$ for $I < I_1$ is positive and grows with increasing $I$ as well,
which is suggestive of the LO effect gradually taking place.
When $I$ is increased further, $V_\mathrm{nl}^{-} (I)$ begins to decay in a way similar to that in Fig.~\ref{hightres}(a), whereas $V_\mathrm{nl}^{+} (I) > 0$ continues to grow until $I=I_2$ is reached, which is a current just after the maximum of $V_\mathrm{nl}(I)$ on the $I >0$ side.
Characteristic currents $I_1$ and $I_2$ are in the inset to Fig.~\ref{intertres}(a) plotted vs $B_\mathrm{ext}$. 
The decrease of $V_\mathrm{nl}^{+} (I)$ after $I$ has exceeded $I_2$ implies a reduction of $f_M$ by $f_T$ that appears due to EH at high $I$. Eventually, $f_T$ prevails and $V_\mathrm{nl}^{+}$ becomes negative but not constant as in Fig.~\ref{rawehlo}(a), which suggests that the superconductivity has survived in the form of a heated LO state. Coexistence of the LO effect and EH was actually predicted theoretically,\cite{lo,bs} but experimental confirmations have been facing difficulties related to weak sensitivity of local $V(I)$ to such subtle effects. 
At $T=2$ K, conditions for this coexistence are just right: $T$ is still close enough to $T_c$ for the quasiparticle distribution function to assume the LO form, but the number of phonons is too small for taking away all the heat if the energy input is large. 

Finally, now it becomes clear why analyses of local $V(I)$ at intermediate $T$ do not give a proper answer on the microscopic mechanisms behind these curves: the SNEQ changes its nature along the $V(I)$.

\subsection{SNEQ regimes in the $T$-$B_\mathrm{ext}$ plane}
\label{btplane}

\begin{figure}
\includegraphics[width=75mm]{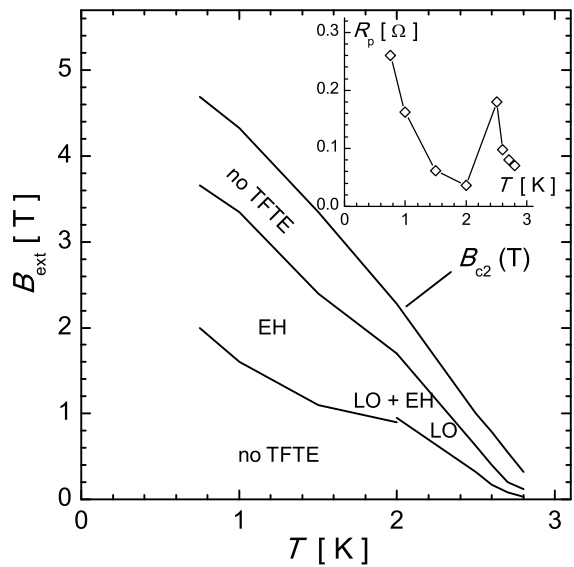}
\caption{Regions of different SNEQ mixed states, as extracted from the TFTE data, plotted in the $T$-$B_\mathrm{ext}$ plane. Inset: $T$ dependence of the maximum nonlocal resistance $R_p$.}
\label{btplanesum}
\end{figure}

We complete our discussion by mapping the TFTE results for the appearance of different SNEQ regimes, which is shown in Fig.~\ref{btplanesum}. The TFTE occurs in a restricted area of the $T$-$B_\mathrm{ext}$ plane. The lower boundary of its appearance is affected mainly by the pinning in the channel, which impedes vortex motion therein and consequently
leads to $V_\mathrm{nl}=0$ when it becomes strong enough at low $T$ and $B_\mathrm{ext}$.
The upper boundary is at the present time less understood. It may reflect a smearing-out of superconducting properties as most of the sample volume becomes normal, so that signatures of some phenomena become
immeasurably small. However, one can also not rule out that it may be associated with high-$b$ fluctuations which in $a$-NbGe films seem to appear in an appreciable $B$ region blow $B_{c2}$.\cite{blatter}
While a full mapping of SNEQ mixed states requires a combination of $V_\mathrm{nl}(I)$ and 
$V_\mathrm{l}(I)$ results, there are situations where $V_\mathrm{l}(I)$ is of little use and 
$V_\mathrm{nl}(I)$ is decisive, for instance in showing that EH and the LO effect 
can coexist at intermediate $T$.

Since $R_\mathrm{nl}$ is also required for understanding and quantifying the TFTE in different regimes,  
in the inset to Fig.~\ref{btplanesum} we show the $T$ dependence of its representative $R_p$ which 
is the maximum of $R_\mathrm{nl}(b)$ extracted from all $V_\mathrm{nl}(I)$ at a given $T$, see Figs.~\ref{lowtres}(b) and \ref{hightres}(b). Actually, $R_p$ is a good estimate for the peak value of $R_\mathrm{nl}(B_\mathrm{ext})$ curve obtained by sweeping $B_\mathrm{ext}$ isothermally at a low $I$, which was the method of Refs.~\onlinecite{grigorieva} and \onlinecite{helzel}.
For $a$-NbGe samples in Ref.~\onlinecite{helzel}, $R_p(T)$ was several times higher than here because these samples had such a low pinning that $\eta \approx \eta_f$ applied close to $T_c$.
However, the shapes of the two $R_p(T)$ curves are very similar. $R_p$ is high at low $T$ because it occurs at high $B_\mathrm{ext}$, and $R_p \propto B_\mathrm{ext}$. There is also an upturn of $R_p$ before the TFTE disappears at $T_c$, because the pinning close to $T_c$ weakens, this reduces $\tilde{\eta}$
and enhances $R_p \propto 1 / \tilde{\eta}$.  This similarity implies that the TFTE does not suffer much from pinning as long as the main effect is in $\eta_f \rightarrow \tilde{\eta}$ due to the shear between vortices 
moving at different $u_\mathrm{nl}$ (which may also include $u_\mathrm{nl}=0$ for some of them).

\section{Summary and conclusions}
\label{conclusions}

In this follow-up to Ref.~\onlinecite{ottoprl}, we present a broader perspective on the transversal flux transformer effect (TFTE) at different local vortex dynamics. At least in weak pinning materials - where fundamental phenomena in vortex motion dominate over sample-dependent pinning - the TFTE is a powerful diagnostic tool for vortex dynamics in the local lead. The TFTE is particularly helpful at high applied currents $I$, where the local mixed-state thermodynamics
is altered. In this case, while the local dissipation curves offer only meager evidence for the microscopic processes being different at low and high temperatures $T$, the TFTE leaves no doubt:
the sign of the nonlocal voltage is opposite in the two cases. This is a consequence of the nonequilibrium quasiparticle distribution function being fundamentally different at low and high $T$, which results in different thermodynamic properties.

At low $T$, the entire quasiparticle system is heated locally. This leads to an expansion of vortex cores, and the corresponding TFTE stems from a $T$ gradient at the interface of the local and nonlocal regions. This Nernst-like effects pushes vortices away from the local region.
Close to $T_c$, the isothermal Larkin-Ovchinnikov effect takes place in the local region, resulting in a shrinkage of vortex cores and an enhanced diamagnetic response. The magnetization gradient at the interface drives vortices toward the local region by a Lorentz-like force. The TFTE at intermediate $T$ shows that the Larkin-Ovchinnikov effect appears at moderate $I$ but it is modified by electron heating at higher $I$, which cannot be concluded from the local current-voltage curves.

Remarkably, these effects - including the TFTE with vortices being locally driven by the Lorentz force - can all be accounted for by a simple model of the magnetic pressure exerted by vortices
under the direct influence of the driving force. The only variable inputs to the model are the type of the driving force and its spatial extent.

The above picture is quantified by an analysis of the nonlocal current-voltage characteristics of a nanostructured $a$-NbGe film, measured over a range of $0.26 \leq T/T_c \leq 0.95$.
The relevant extracted quantities are the nonlocal resistance in the low-$I$ linear response regime, the vortex transport entropy of the Nernst-like effect at low $T$, and the magnetization enhancement at $T \lesssim T_c$ together with the consequent interface current that produces the local Lorentz-like force.

\begin{acknowledgments}
This work was supported by the DFG within GK 638. A.~B. acknowledges support from the Croatian Science
Foundation (NZZ), D.~Yu.~V. from Dynasty Foundation, and D.~B. from Croatian
MZOS project No. 119-1191458-1008.

\end{acknowledgments}

\appendix

\section{Enhancement of the mixed-state diamagnetism by the LO effect}
\label{LOmag}

In order to find $\delta M$, we calculate, along the LO formalism,\cite{lo} the magnetic moment ${\bf m}=(1/2)\int [{\bf r} \times {\bf j}_s]dS_\mathrm{cell}$ of a single-vortex cell, where the supercurrent around the vortex core is given by
\begin{equation}
\label{supercurr}
{\bf j}_s=\frac{1}{\rho_\mathrm{n} e}\left(\frac{\pi}{4k_\mathrm{B}T_c}|\Delta|^2+\frac{\pi}{2} |\Delta|\delta
g(|\Delta|)\right)\left(\nabla \varphi-\frac{2e}{\hbar}{\bf A}\right) \; ,
\end{equation}
{\bf A} being the vector potential and $\Delta=|\Delta|\mathrm{exp}(i\varphi)$ the order parameter.
$\delta g (\epsilon)$ is the nonequilibrium correction to the equlibrium quasiparticle distribution function $g_\mathrm{eq}(\epsilon)=\tanh(\epsilon/2k_\mathrm{B}T)$ for quasipartices of energy $\epsilon$, so that the nonequilibrium distribution function is $g_\mathrm{neq} = g_\mathrm{eq} + \delta g$. The dipole magnetic field created by {\bf m} of a given cell opposes ${\bf B}_\mathrm{ext}$ in the surrounding cells and thus enhances the diamagnetic response. By setting $\delta g = 0$ and $\delta g \neq 0$ in Eq.~(\ref{supercurr}) for the equilibrium and nonequilibrium situations, respectively, and by
summing-up the dipole field over the entire lattice, we can find $M_\mathrm{eq}$ and $M_\mathrm{neq}$. In the calculation, the Wigner-Seitz cell of the Abrikosov lattice is replaced by a
circle of a radius $r_B=\sqrt{\phi_0/\pi B}$.

We have to find $\Delta$ and $\delta g$ in order to calculate  ${\bf j}_s$. Since $T$ is close to $T_c$, we can use the modified Ginzburg-Landau equation
\begin{eqnarray}
\label{modGL}
|\Delta|\left[ 1-|\Delta|^2-(1/r-Br/2)^2+\Phi_1 \right] +
\\
\nonumber 
+\frac{1}{r}\frac{\partial}{\partial r} \left( r
\frac{\partial |\Delta|}{\partial r} \right)
 +\frac{1}{r^2}\frac{\partial^2 |\Delta|}{\partial
\alpha^2} =0 \; ,
\end{eqnarray}
where $(r,\alpha)$ defines the two-dimensional polar coordinate system. Here and below, we use dimensionless units. The order parameter and energy are in units of $\Delta_0(T)=\sqrt{8\pi^2/7\zeta(3)} k_\mathrm{B}T_c\sqrt{1-T/T_c}\simeq 3.06 k_\mathrm{B}T_c\sqrt{1-T/T_c}$ [where $\zeta$ is the Riemann's zeta
function], length is in units of $\xi(T)=\sqrt{\pi \hbar D/8 k_\mathrm{B} (T_c-T)}$, and magnetic field is in units of $B_{c2}(T)=\phi_0/2\pi\xi^2(T)$.
\begin{equation}
\label{Phi_1}
\Phi_1=\frac{1}{1-T/T_c}\int_{\Delta}^{\infty} \frac{\delta
g(\epsilon)  d \epsilon }{(\epsilon^2-|\Delta|^2)^{1/2}}
\end{equation}
describes the influence of $\delta g$. The boundary conditions in Eq.~(\ref{modGL}) are $|\Delta|_{r=0}=0$ and [$\partial |\Delta|/\partial r]_{r=r_B}=0$. Eq.~(\ref{modGL}) for $|\Delta|$
is coupled with the following Boltzmann-like equation:
\begin{eqnarray}
\label{Boltz}
\nonumber
-\frac{1}{r}\frac{\partial}{\partial r} \left(r \frac{\partial
\delta g}{\partial r} \right)-\frac{1}{r^2}\frac{\partial^2 \delta
g} {\partial \alpha^2}+\frac{\partial \delta g}{\partial
\tau}\frac{\partial (\epsilon^2-|\Delta|^2)^{1/2}}{\partial
\epsilon}-
\\ 
\\
\nonumber
-\frac{\partial (g_0+\delta g)}{\partial \epsilon}\frac{\partial
(\epsilon^2-|\Delta|^2)^{1/2}}{\partial \tau}=-\frac{\delta
g}{L_\Sigma^2} \frac{|\epsilon|}{(\epsilon^2-|\Delta|^2)^{1/2}} \; ,
\end{eqnarray}
where $\tau$ denotes time (in units of $\tau_0 = \xi^2/D$), and $L_\Sigma = \sqrt{D \tau_\mathrm{e,ph}} / \xi$ is a dimensionless inelastic electron-phonon  relaxation length. The above equation is valid for $|\epsilon| > |\Delta|(r)$ and $\delta g \ll g_\mathrm{eq}$. It can be simplified for $B_{c1}
\ll B \sim B_{c2}$ (i.e., $r_B / \xi \ll L_\Sigma$), where $B_{c1}$ is the lower critical magnetic field, and a relatively weak electric field $E$ (see Ref.~\onlinecite{lo}). In this case, one can seek for its solution in the form $\delta g = \langle g \rangle + g_1 (r,\alpha)$, where $g_1$ is proportional to vortex velocity $u$, and the coordinate-independent term $\langle g \rangle$ is proportional to $u^2$. The natural scale
for $u$ in our units is $u_0=\xi/\tau_0$ (below we also use the expression for the LO velocity
$u_\mathrm{LO}=[14 D \zeta(3)(1-T/T_c)^{1/2}/\pi \tau_\mathrm{e,ph}]^{1/2}$). In
this limit, the equation for $g_1$ is given by
\begin{equation}
\label{g1}
\frac{1}{r}\frac{\partial}{\partial r} \left(r \frac{\partial
g_1}{\partial r} \right)+\frac{1}{r^2}\frac{\partial^2
g_1}{\partial \alpha^2}=\frac{\partial g_0}{\partial \epsilon
}\frac{u}{u_0} \cos (\alpha) \frac{\partial
(\epsilon^2-|\Delta|^2)^{1/2}}{\partial r} \; ,
\end{equation}
The main effect on $|\Delta|$ arises from $\langle g \rangle$. For
that reason, one can also neglect the angular dependence of
$|\Delta|$ in Eq.~(\ref{modGL}). By solving Eq.~(\ref{g1})
(with a boundary condition $\partial g_1/\partial r = 0$ at $r = r_\epsilon$ and $r = r_B$),
inserting the result into Eq.~(\ref{Boltz}) and averaging it over coordinates, one
obtains
\begin{equation}
\label{gaverage2}
\langle g \rangle = \left(\frac{u}{u_\mathrm{LO}} \right)^2
\left(1-\frac{T}{T_c} \right) \frac{\partial {\cal D}
(\epsilon)}{\partial \epsilon} \frac{1}{{\cal D}_1(\epsilon)} \; ,
\end{equation}
where
\begin{equation}
\label{D1}
{\cal D}_1(\epsilon)=|\epsilon| \int_0^{r_B}\frac{r
dr}{(\epsilon^2-|\Delta|^2)^{1/2}} \; ,
\end{equation}
\begin{eqnarray}
\label{D}
\nonumber
{\cal D}(\epsilon)=\int_0^{r_B} dr \frac{\partial
(\epsilon^2-|\Delta|^2)^{1/2}}{\partial r} \times
\\
\times \int_0^{r} dr_1 r_1
[(\epsilon^2-|\Delta|^2)^{1/2}+C(\epsilon)],
\end{eqnarray}
with
\begin{eqnarray}
\label{C1}
C(\epsilon)=\frac{2}{r^2_\epsilon}\int_0^{r_\epsilon} r dr
(\epsilon^2-|\Delta|^2)^{1/2}
\end{eqnarray}
for  $|\epsilon|<|\Delta|_\mathrm{max}$, and
\begin{eqnarray}
\label{C2}
C(\epsilon)=-2(\epsilon^2-|\Delta|_\mathrm{max})^{1/2}+
\\ \nonumber
+\frac{2}{r^2_B}\int_0^{r_B} r dr  (\epsilon^2-|\Delta|^2)^{1/2},
\end{eqnarray}
for $|\epsilon| > |\Delta|_\mathrm{max}$. Equating $|\Delta| (r_\epsilon) = \epsilon$ gives $r_\epsilon$, and, with that, the set of Eqs.~(\ref{modGL},\ref{gaverage2}-\ref{C2}) is approached
numerically.

In Ref.~\onlinecite{ottoprl}, the numerical calculation was is carried out for $T=0.85 T_c$ and $0.2<B/B_{c2}<1$, with a restriction to $0\leq u\leq 2u_\mathrm{LO}$. Namely, at this $T/T_c$, the
LO approach becomes inapplicable above $u \simeq 2u_\mathrm{LO}$ because $\xi$ approaches $\xi(0)$ and the local approximation for normal and anomalous Green's functions cannot be used. The
calculation results showed that $\langle g \rangle$ was positive at $\epsilon < |\Delta|_\mathrm{max}$, and negative at $\epsilon > |\Delta|_\mathrm{max}$, which resulted in $\Phi_1 >0$  near the
vortex core (leading to an enhancement of $|\Delta|$ and a shrinking of the core) and $\Phi_1 <0$ far away from the vortex core (leading to a suppression of $|\Delta|$ there). Application of
this to Eq.~(\ref{supercurr}) and consequent calculation of $M$, as explained before, led to $|M_\mathrm{neq}| > |M_\mathrm{eq}|$, see Fig.~3 of Ref.~\onlinecite{ottoprl}.

\section{Influence of quasiparticle diffusion on the LO effect in the local lead}
\label{qpdiff}

\begin{figure}
\includegraphics[width=75mm]{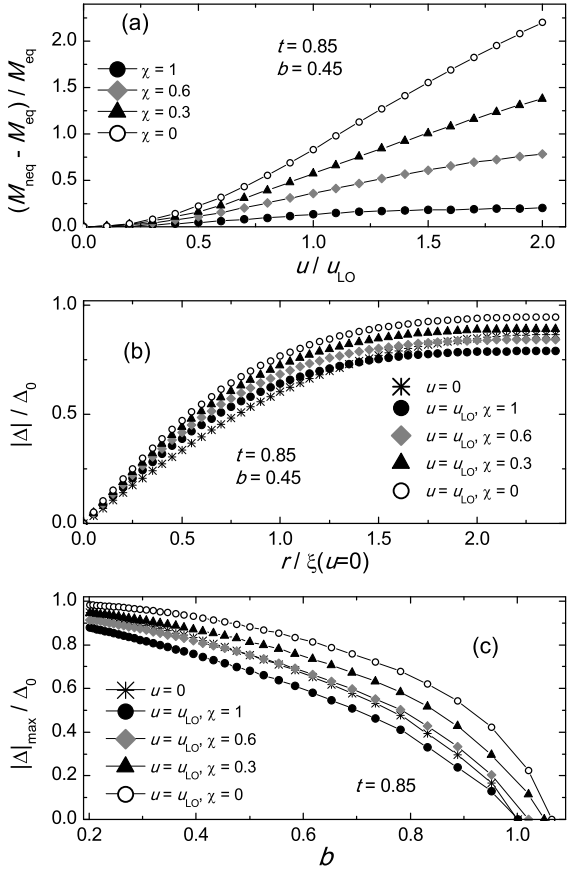}
\caption{Calculated influence of $\chi$ on the LO effect in the local lead at $t=0.85$ and for $\chi=0,0.3,0.6,1$. (a) Diamagnetism enhancement at $b=0.45$, as a function of $u/u_\mathrm{LO}$. (b) $|\Delta|/\Delta_0$ against
$r/\xi(u=0)$ in the single-vortex cell for $b=0.45$ and $u = u_\mathrm{LO}$.
(c) $|\Delta|_\mathrm{max}/\Delta_0$ vs $b$, at $u = u_\mathrm{LO}$.}
\label{theory}
\end{figure}

The calculation in Appendix~\ref{LOmag} assumes an infinite vortex
lattice. However, in our experiment, the LO effect occurs in the
$W$-wide section of the local lead, see Fig.~\ref{sample}(a),
whereas in the rest of the sample  the vortex lattice essentially
preserves its equilibrium properties. At $T=2.5$ K, 
$L_\varepsilon \sim 270$ nm is comparable to the length of the
$W$-wide section of the local lead,\cite{ottodiss} so majority
of the quasiparticles with $\epsilon > |\Delta|_\mathrm{max}$ can
diffuse into the adjacent areas where there is
no LO effect. Since $\delta g <0$ for $\epsilon >
|\Delta|_\mathrm{max}$, the removal of these quasiparticles
should further enhance $|\Delta|$ near the cores and consequently also the diamagnetism in the local lead
[the nonequilibrium contribution $\Phi_1$ given by Eq.~(\ref{Phi_1}) is in this case larger and 
contributes more strongly to Eq.~(\ref{modGL})]. This
anticipation should be taken into account in future sample design
for experiments relying on the TFTE, i.e., $\delta M$ is expected to
be smaller if the narrow part of the local lead is longer.

In order to estimate the effect of the diffusion, one would have
to solve the equation for $\delta g$ for the whole sample, which
is rather intricate. However, there is a simpler approach which
can provide ample information as well: we parametrize the
quasiparticle removal efficiency by multiplying $\delta g$ by a
factor $0 \leq \chi \leq 1$ for all quasiparticles with $\epsilon
> |\Delta|_\mathrm{max}$. Physically, $\chi=0$ corresponds to
complete removal of these quasiparticles from the local lead,
and $\chi=1$ to no removal at all. For our sample, we estimate $\chi
\approx 0.6$ on the basis of solving a two-dimensional diffusion
equation for $\delta g$ at $\epsilon = |\Delta|_\mathrm{max}$,
with uniformly distributed fourth term in Eq.~(\ref{Boltz}) in the
region where the LO effect takes place.

Results of carrying out the calculation in the same way as as in
Appendix~\ref{LOmag} but with $\delta g \rightarrow \chi \delta g$
are displayed in Fig.~\ref{theory}, for $t=0.85$ and $\chi=0,0.3,0.6,1$. 
In Fig.~\ref{theory}(a), we show 
$(M_\mathrm{neq} - M_\mathrm{eq})/M_\mathrm{eq}$ at $b=0.45$ as a function of
$u/u_\mathrm{LO}$. A monotonic increase of the magnetization
enhancement as $\chi$ decreases, i.e., as the removal efficiency
grows, is discernible.\cite{chi} In Fig.~\ref{theory}(b), we plot $|\Delta|
/ \Delta_0$ in and around the vortex core, again for $b=0.45$, vs
$r$ scaled to the coherence length $\xi(u=0)$ at zero vortex
velocity. Results are shown for  static vortices ($u=0$) and for
vortices moving at $u=u_\mathrm{LO}$. It can be seen that the
gap enhancement is stronger for smaller $\chi$. In Fig.~\ref{theory}(c), we
plot $|\Delta|_\mathrm{max} / \Delta_0$ vs $b$ for
$u=u_\mathrm{LO}$, where it is visible that
$|\Delta|_\mathrm{max}$ is also enhanced when $\chi$ decreases.
Interestingly, the model predicts a survival of superconductivity
at $b > 1$ for low $\chi$. The effect is similar to the
enhancement of the critical current and critical temperature,
induced by microwave radiation.\cite{dgn,ei} The difference
is in the source of nonequilibrium, 
which for a rapidly moving vortex lattice is the time variation of $\Delta$.


\begin{thebibliography}{}
\bibitem{errbars} Error bars in all plots where we use symbols are for clarity not shown because they do
                not exceed the symbol size much.
\bibitem{superj} In that sense, the strict expression for $f_\mathrm{L}$ is obtained with
                $j \rightarrow j_s$, which is for simplicity not pursued throughout this paper.
\bibitem{chi} Effect of $\chi$ was not taken into account in Ref.~\onlinecite{ottoprl} where                              			 		$\chi=1$ was assumed in the calculation. Note that the stronger enhancement of 
							$\delta M$ for $\chi = 0.6$ leads to a better quantitative agreement of calculated $\delta M$
							with that shown in Fig.~\ref{hightres}(c).
\end{thebibliography}
\end{document}